\documentclass[12pt]{article}
\usepackage{amsmath,amssymb}
\RequirePackage[cp1251]{inputenc}
\usepackage{graphicx}
\begin{document}

 \begin{center}

{\large\bf  Optically induced electrostriction modes in a nanoparticle of a uniformly
charged electret}

 \vspace*{0.5cm}

S.I. Bastrukov\footnote{Corresponding author,
 e-mail: bast@phys.nthu.edu.tw}$^{,2}$, I.V. Molodtsova$^2$, Pik-Yin Lai$^{3}$

 \vspace*{0.5cm}

{\it $^1$ Department of Physics and Institute of Astronomy, \\
 National Tsing Hua University, Hsinchu, 30013 Taiwan\\

$^2$ Laboratory of Informational Technologies,\\
 Joint Institute for Nuclear Research, 141980 Dubna, Russia\\

$^3$ Department of Physics and Center of Complex Systems, \\
National Central University, Chung-li,  320 Taiwan

}

 \end{center}

\begin{abstract}
 The electromagnetic response of a nanoparticle of an ion-doped polymeric elastic insulator, commonly called as an electret, is considered in the continuum model of a uniformly charged elastic sphere. The spectral formulae for the frequency of optically induced spheroidal and torsional shear oscillations driven by bulk force of elastic and dielectric stresses are obtained in analytic form. Particular attention is given to relaxation dielectric mode of the electrostriction response and its stability in the lowest quadrupole mode. The practical
 usefulness of ultrafine particles of electrets as a biolabels capable of accumulating likely-charged inclusions uniformly dispersed over the spherical volume of an elastic matrix is briefly discussed.
\end{abstract}

\vspace*{0.1cm}


\vspace*{0.5cm}

\section{Introduction}
 The phenomenological continuum-mechanical models of vibrating liquid drop
 and solid globe provide proper account of gross features of resonance-like excitations in electromagnetic
 response of ultra fine particles. The spectra of these excitations are currently measured by methods of Raman scattering and inelastically
 scattered neutrons. The physical idea underlying these models is to identify the frequency of ac field incident on a nanoparticle with frequency of material oscillations in the particle regarded as a spherical piece of
 either liquid or solid continuous medium.
 A canonical example is fluid-mechanically computed Mie's spectral formula for the frequency of surface
 plasmons in nanoparticles of highly conducting metals like silver and gold [1-5].
 Another example is the solid-mechanically calculated
 frequency spectra of optically induced elastic oscillations, acoustic phonons, in a nanoparticle of a dielectric solid [6-8].

 In this work we consider a model of resonant response of a nanoparticle of charged electret
 (insulating elastic matrix doped by likely-charged ions homogeneously distributed over spherical volume)
 responding to electromagnetic load by shear electro-mechanical fluctuations which are described
 in the approximation of a uniformly charged elastic continuum.
 To illuminate the difference between the electric reactive forces driven by optically-induced electro-mechanical
 vibrations in a nanoparticle of a uniformly charged electret (dielectric material) and plasma
 oscillations of free electrons against immobilized ions in a nanoparticle of
 noble metals (highly conducting materials), in Sec.2 a brief outline is given of classical electron theory of the surface plasmons and solid-mechanical theory of acoustic phonons. In Sec.3, the frequency spectra of electrostriction spheroidal and torsional vibrational modes in the ion-doped dielectric
 nanoparticle of electret are obtained followed by discussion of electro-elastic stability of such a nanoparticle
 to optically induced distortions, the issue of crucial importance to their use as biolabels. The results are briefly summarized in Sec.4.

 \section{Resonance frequencies of surface plasmons and acoustic phonons}

 Within the framework of  the classical electron
 theory of metals, the low-frequency
 electromagnetic response of metallic nanoparticle to long-wavelength ac field
 is modeled by oscillations of electron liquid against crystal lattice treated as a static uniformly charged ionic background and described by coupled equations of electron fluid-mechanics and ionic
 electrostatics
 \begin{eqnarray}
 \label{e2.1}
 && \rho\frac{\partial \delta {\bf v}}{\partial t}=
 \rho_e\delta {\bf E},
 \quad\quad\rho=m_e^*n_e,\quad\rho_e=-en_e.
  \end{eqnarray}
 The emergence of Coulomb restoring force is attributed to
 a small amplitude fluctuations of intrinsic electric field $\delta {\bf E}$ (between free
 electrons and ions) superimposed on electrostatic field of ionic jellium,
 ${\bf E}_0$ which is computed from the equation $\nabla\cdot {\bf E}_0=4\pi
 \rho_i$, where $\rho_i=en_i$; the particle number of electrons $n_e$
 equals to that of ions $n_i$ due to electro-neutrality of metal nanoparticle.
 The effective mass of electron $m_e^*$ is unique to each specific metal.
 The effect of resonance photo-absorption emerges when the frequency of incident on particle ac
 field $\delta {\bf E}$ becomes equal to the
 frequency of oscillations of the velocity $\delta {\bf v}$
 of electron flow obeying the vector Laplace equation:
 \begin{eqnarray}
 \label{e2.2}
 \nabla^2 \delta {\bf E}=0,\quad\quad  \nabla^2 \delta {\bf v}=0.
 \end{eqnarray}
 These equations can be thought of as long wavelength limit of the Helmholtz equation for standing
 waves and describing electromagnetic response by long wavelength vibrations
 of electric field and electron flow, Rayleigh regime.
 In this regime the frequency of oscillations of free electrons against ionic background
 can be computed by the energy method which is based on the equation of energy balance
 \begin{eqnarray}
 \label{2.3}
 && \frac{\partial }{\partial t}
 \int \frac{\rho \delta{\bf v}^2}{2}\,d{\cal V}=
 \int \rho_e (\delta {\bf v}\cdot\delta {\bf E})\, d{\cal V}.
  \end{eqnarray}
 The frequency of surface plasmons as a function of multipole degree $\ell$ is given by the well-known
 Mie spectral formula [1-4]:
 \begin{eqnarray}
 \label{e2.4}
 \omega^2_{\rm Mie}(\ell)=\omega^2_p\frac{\ell}{2\ell+1}\quad\quad
 \omega^2_p=\frac{4\pi n_e e^2}{m_e^*}\quad\quad \ell\geq 1.
 \end{eqnarray}
 The dipole surface plasmon resonance whose energy is given by, $E_{\rm Mie}(\ell=1)=\hbar\omega_p/\sqrt{3}$,
 is well singled out in the cross sections of photo-absorption by silver and gold
 nanoparticles.

 Also relevant to our further discussion is the continuum, solid-mechanical, description
 of acoustic phonons, quanta of low-frequency elastic oscillations of solid.
 The mathematical treatment of such
 vibrations rests on canonical equation of solid-mechanics for the field of material distortions $u_i({\bf r},t)$
\begin{eqnarray}
 \label{e2.5}
&& \rho{\ddot u}_i=\nabla_k\sigma_{ik},\quad\sigma_{ik}=2\,\mu\, u_{ik},\quad u_{ik}=\frac{1}{2}[\nabla_i u_k+\nabla_k u_i],\quad
u_{kk}=\nabla_k\,u_k=0.
 \end{eqnarray}
 where $\rho$ stands for the density the shear modulus $\mu$
 relates applied shear
 stresses  $\sigma_{ik}$ and resulting shear or strains $u_{ik}$ in accord with the law of Hooke.
 The conservation of energy is controlled by equation
\begin{eqnarray}
 \label{e2.6}
  \frac{\partial }{\partial t}\int \frac{\rho {\dot u}^2}{2}\,d{\cal
  V} = -\int \sigma _{ik}{\dot u}_{ik}\,d{\cal V},\quad\quad {\dot
 u}_{ik}=\frac{1}{2}[\nabla_i\,{\dot u}_k+\nabla_k\,{\dot u_i}].
   \end{eqnarray}
 The fluctuating fields of material displacements $u_i({\bf r},t)$ and strains $u_{ik}({\bf r},t)$
 can be conveniently represented in the following separable form
 \begin{eqnarray}
 \label{e2.7}
 {\bf u}({\bf r},t)={\bf a}({\bf r})\,{\alpha}(t),\quad u_{ik}({\bf r},t)=a_{ik}({\bf r})\alpha(t),\quad
 a_{ik}=\frac{1}{2}[\nabla_i\,a_k+\nabla_k\,a_i]
 \end{eqnarray}
 with help of which the equation of energy balance (\ref{e2.6})
 is reduced to equation for $\alpha(t)$ having the form of standard equation of linear oscillations:
 \begin{eqnarray}
\label{e2.8}
 M{\ddot \alpha}(t)+K\alpha(t)=0,\quad M=\int \rho\, a_i({\bf r})\,a_i({\bf r}) d{\cal V},\quad
  K=2\int \mu\, a_{ik}({\bf r})\,a_{ik}({\bf r}) d{\cal V}.
 \end{eqnarray}
 In Rayleigh's regime, the vibrational response of an elastic sphere
 is characterized by two fundamental modes of shear elastic oscillations
 of the nodeless fields of material displacements which are specified by
 two fundamental solutions of the vector Laplace equation
  \begin{eqnarray}
 \label{e2.9}
 \nabla^2  {\bf u}({\bf r},t)=0\quad\to\quad  \nabla^2 {\bf a}({\bf r})=0.
 \end{eqnarray}
 The instantaneous displacements in spheroidal mode of irrotational oscillations -- ${\bf a}_s$
 is given by poloidal field and in torsional mode of differentially rotational oscillations -- ${\bf a}_t$
 by the toroidal fields, respectively \cite{B-07}:
 \begin{eqnarray}
 \label{e2.10}
&& {\bf a}_s=\nabla \times \nabla\times ({\bf
 r}\,\chi),\quad\quad {\bf a}_t=\nabla \times  ({\bf
 r}\chi),\quad
 \chi({\bf r})={\cal N}_\ell\, r^\ell\,P_\ell(\cos\theta).
 \end{eqnarray}
 Henceforth $P_\ell(\cos\theta)$ stands for the Legendre polynomial of multipole
 degree $\ell$ and ${\cal N}_\ell$ is the arbitrary constant.
 The spectral equations for the frequencies of the even parity spheroidal
 $\omega_s$ and the odd parity torsional $\omega_t$ modes of acoustic phonons are
 given by \cite{B-94,B-07}:
 \begin{eqnarray}
 \label{e2.11}
  && \omega_s(\ell)=\omega_0[2(2\ell+1)(\ell-1)]^{1/2},\quad\quad \omega_t(\ell)=\omega_0[(2\ell+3)(\ell-1)]^{1/2},\quad\ell\geq 2,\\
   \label{e2.12}
  && \omega_0=\frac{c_t}{R},\quad c_t=\sqrt{\frac{\mu}{\rho}}
 \end{eqnarray}
where $c_t$ is the speed of elastic shear wave in the
material bulk and $R$ stands for the radius of nanoparticle. These last
spectral equations are central to spectroscopic analysis of optical
modes excited by inelastically scattered x-rays and neutrons.
The purpose of the above outlined models
 is to illuminate similarity and difference between spectral equations for the frequencies of surface plasmons, equation (\ref{e2.5}), and the frequency spectra of electrostriction modes in a nanoparticle of a uniformly charged electret which is the main subject of the next section.

 \section{Electrostriction mode in optical response of nanoparticle of a uniformly charged electret}

 The optically-induced, by ac electromagnetic field, electro-mechanical distortions in the volume
 of uniformly charged electret are characterized by the tensor of dielectric stresses (e.g., \cite{Ho}):
  \begin{eqnarray}
 \label{e3.1}
 && p_{ik}=\frac{1}{8\pi} [E_i\,\delta D_k+E_k\,\delta D_i-(E_j\,\delta D_j)\delta_{ik}]
 \end{eqnarray}
 where $E_i$ are components of electrostatic field produced in the particle volume by
 extraneous charge uniformly distributed with the charge density $\rho_e$:
 \begin{eqnarray}
 \label{e3.2}
 \nabla\cdot {\bf E}=4\pi\frac{\rho_e}{\epsilon},\quad \nabla\times{\bf E}=0
 \end{eqnarray}
 whose solution is well known:
 \begin{eqnarray}
 \label{e3.2a}
 {\bf E}({\bf r})=-\nabla \Phi({\bf r}),\quad
 \Phi(r)=-\frac{2\pi}{3\epsilon}\rho_e(r^2-3\,R^2),\quad
 [E_r=\frac{4\pi}{3}\,\frac{\rho_e}{\epsilon}\,r,\quad
 E_{\theta}=0,\quad E_{\phi}=0].
 \end{eqnarray}
  The electro-mechanical effect is described by constituting equation\footnote{
 The constitutive equation for $\delta D_i$ is compatible with
 the Maxwell equation
 $\nabla\times \delta {\bf H}=(4\pi/c)\delta {\bf j}+(1/c)\delta {\dot {\bf D}}$.
 Applying to this latter equation operator of divergence and taking into account that $\delta {\bf j}=\rho_e\delta {\bf v}$ and the continuity equation of the charge conservation $\delta {\dot \rho}_e=-\nabla\,\delta {\bf j}$,
 we obtain $\delta {\dot {\bf D}}=-4\pi\rho_e\delta {\bf v}$. Bearing in mind that $\delta {\bf v}={\dot {\bf
 u}}$ and eliminating in the last equation the time derivative one arrives at (\ref{e3.3}).}
 \begin{eqnarray}
 \label{e3.3}
 &&\delta D_i=-4\pi\rho_e\,u_i
 \end{eqnarray}
 showing that optically induced fluctuations of dielectric induction $\delta D_i$ are linearly proportional
 to material displacements $u_i$ and inextricably related
 to the storage of extraneous charge uniformly dispersed with density $\rho_e$ over the sample volume;
 note, the dielectric materials can accommodate only extraneous charge.

\subsection{Dielectric relaxation mode}

 We confine our analysis to the Rayleigh's regime of optical perturbation
 resulting in non-compressional fluctuations of the electret material
 (the charge density remains unchanged $\delta \rho_e=-\rho_e\,\nabla_k\,u_k=0$)
 which are described by nodeless field of material displacements $u_i$ obeying the vector Laplace equation, $\nabla^2 {\bf u}=0$. The purpose of this subsection is to elucidate the difference between plasma oscillations of free electrons against ionic jellium in a metal nanoparticle which are restored by electrostatic
 force of the form $f_i=\rho_e\,E_i$, Eq.(\ref{e2.1}),
 and the shear oscillations in a nanoparticle of uniformly charged electret which are restored by
 reactive force of dielectric stresses $f_i'=\nabla_k\,p_{ik}$ which are governed by
 equations
 \begin{eqnarray}
 \label{e3.4}
 \rho\,{\ddot u}_i=\nabla_k p_{ik},\quad
 \frac{\partial }{\partial t}\int \frac{\rho {\dot u}^2}{2}\,d{\cal V}=
 -\int p_{ij}\,{\dot u}_{ij}\,d{\cal V},\,\, {\dot
 u}_{ik}=\frac{1}{2}[\nabla_i\,{\dot u}_k+\nabla_k\,{\dot
 u_i}],\,\, u_{kk}=0.
 \end{eqnarray}
 The rightmost equation shows  that optically induced dielectric stresses $p_{ik}$ are accompanied by shear deformations $u_{ik}$. The eigenfrequency can be
 computed with help of the above expounded Rayleigh's energy method whose key point is
 the separable representation of fluctuating material displacements and strains in the form
 given by equations (\ref{e2.7}). Then, for the perturbation-induced dielectric induction
 one has
 \begin{eqnarray}
 \label{e3.5}
\delta D_i({\bf r},t)=-4\pi\rho_e\,a_i({\bf r})\alpha(t)
\end{eqnarray}
 and the tensor of dielectric stress is given by
 \begin{eqnarray}
 \label{e3.6}
 && p_{ik}({\bf r},t)=[\tau_{ik}({\bf r})-\frac{1}{2}\tau_{jj}\delta_{ik}]\alpha(t),\quad \tau_{ik}({\bf r})=-\rho_e
 [E_i({\bf r})\,a_k({\bf r})+E_k({\bf r})\,a_i({\bf r})].
 \end{eqnarray}
 Substitution (\ref{e2.7}), (\ref{e3.5}) and (\ref{e3.6}) into equation of energy balance in (\ref{e3.4}) leads
 to
 \begin{eqnarray}
 \label{e3.7}
 &&\frac{dH}{dt}=0\quad H=\frac{M{\dot \alpha}^2}{2}+
 \frac{K_d{\alpha}^2}{2}\quad\to\quad {\ddot \alpha}(t)+\omega^2\alpha(t)=0,\quad \omega^2=\frac{K_d}{M}\\[0.2cm]
 \label{e3.8}
 && M=\int \rho a_i\,a_i d{\cal V},\quad K_d=\int \tau_{ik}({\bf r})a_{ik}({\bf r}) d{\cal V}\\
 \nonumber
 && \tau_{ik}({\bf r})=-\rho_e
 [E_i({\bf r})\,a_k({\bf r})+E_k({\bf r})\,a_i({\bf r})],\quad
 a_{ik}=\frac{1}{2}[\nabla_i\,a_k+\nabla_k\,a_i].
 \end{eqnarray}
 The fact that such response is accompanied by internal shear
 deformations suggests that dielectric modes in question can be specified in a manner of vibrational
 modes in an elastic sphere, that is, as spheroidal and torsional ones.

 {\it Dielectric spheroidal mode.}
 In this positive parity mode the displacements are described by the poloidal (polar)
 vector field
 \begin{eqnarray}
 \label{e3.10}
 {\bf a}_s={\cal N}_{\ell}\nabla \times \nabla\times {\bf r}\,r^\ell\,P_\ell(\cos\theta).
 \end{eqnarray}
 For the inertia, stiffness and frequency $\omega_{ds}$ of spheroidal dielectric mode we obtain
 \begin{eqnarray}
 \nonumber
&& M^s=4\pi\rho {\cal N}_{\ell}^2 R^{2\ell+1}\frac{\ell(\ell+1)^2}{(2\ell+1)},\quad
 K_d^s=-\frac{32\pi^2}{3}\frac{\rho_e^2}{\epsilon} {\cal N}_{\ell}^2 R^{2\ell+1}\frac{\ell(\ell+1)(\ell^2-1)}
  {(2\ell+1)},
 \\[0.2cm]
 \label{e3.12}
&& \omega_{ds}^2(\ell)=-\frac{2}{3}\omega_d^2\,(\ell-1),\quad
 \omega_d^2=\frac{4\pi}{\epsilon}\frac{\rho_e^2}{\rho}
 \end{eqnarray}
 where $\omega_d$ is the natural unit of dielectric
 fluctuations.

 {\it Dielectric toroidal mode.}
 The material displacements in negative parity torsional mode are described by the axial toroidal (axial)
 vector field
\begin{eqnarray}
 \label{e3.13}
 {\bf a}_t={\cal N}_{\ell}\nabla\times {\bf r}\,r^\ell\,P_\ell(\cos\theta).
 \end{eqnarray}
 Computation of the the inertia, stiffness and frequency $\omega_{dt}$ of torsional dielectric  mode yields
 \begin{eqnarray}
 \nonumber
 && M^t=4\pi\rho {\cal N}_\ell^2 R^{2\ell+3}\frac{\ell(\ell+1)}{(2\ell+1)(2\ell+3)},\quad
 K_d^t=
 -\frac{16\pi^2}{3}\frac{\rho_e^2}{\epsilon} {\cal N}_\ell^2 R^{2\ell+3}\,
 \frac{\ell(\ell^2-1)}{(2\ell+1)(2\ell+3)},\\[0.2cm]
 \label{e3.14}
 && \omega_{dt}^2(\ell)=\frac{K_t}{M_t}=-\frac{1}{3}\omega_d^2\,(\ell-1),\quad
 \omega_d^2=\frac{4\pi}{\epsilon}\frac{\rho_e^2}{\rho}.
 \end{eqnarray}
 The obtained spectral equations represent one of the main newly obtained result of the
 presented theory showing that optical response of nanoparticle of electret are characterized by two
 different in parity modes, even-parity spheroidal dielectric mode and odd-parity torsion one.
 The basic {\it dielectric} frequency
 $\omega_d$ depends upon the dielectric constant $\epsilon$ in such
 a way that in the limit $\epsilon\to \infty$, as is the case of conductors,
 $\omega_d\to 0$. This means that, contrary to the surface plasmons in a nanoparticle of a highly conducting noble metals, the electrostatic fluctuations in a nanoparticle of a uniformly charged electret are manifested as
 relaxation modes, not oscillatory. It is relevant to note that the somewhat different method
  of computing the frequency of nodeless electro-elastic vibrations of a uniformly charged dielectric globe
  that has been developed in \cite{BP-98,WS-99} leads to identical analytic estimates for the frequencies
  of considered dielectric modes.
  The negative sign of squared frequencies, highlighting relaxation nature of dielectric mode, indicates
  that nanoparticle is unstable to optically induced electro-mechanical distortions\footnote{The origin of this instability can be traced from equations
 of electro-elastic solid mechanics describing propagation of perturbations
 in bulk of uniformly charged electret with constant shear modulus $\mu$ which is given by
 $$\rho\,{\ddot u}_i=\nabla_k
 \sigma_{ik}+\nabla_i\,p_{ik}=\mu\,\nabla^2\,{\bf u} +
 ({4{\pi}}/{\epsilon})\,\rho_e^2\,{\bf u}.$$
 For the plane-wave form of externally induced displacements,
 ${\bf u}={\bf u}^0_t
 \exp{i({\bf k}{\bf r}-\omega
 t)}$, the last equation leads to dispersion equation
 \begin{eqnarray}
 \nonumber
 \omega^2=c_t^2\,k^2-\omega_d^2 ,\quad\quad
 c_t^2=\frac{\mu}{\rho},\quad\quad \omega_d^2=\frac{4\pi}{\epsilon}\frac{\rho_e^2}{\rho}
 \end{eqnarray}
 where $c_t$ is the speed of transverse shear wave in the bulk electret
 and $\omega_d$ stands for the natural unit of frequency of dielectric
 oscillations. It follows that there is critical wave number $k_c$ (and corresponding wavelength $\lambda_c=2\pi/k_c$ such that the electret material becomes electro-mechanically unstable when
\begin{eqnarray}
 \nonumber
 k < k_{c}=\sqrt{\frac{4\pi\rho_e^2}{\epsilon\mu}},\quad\quad
  \lambda > \lambda_{c}=\sqrt{\frac{\pi\epsilon\mu}{\rho_e^2}}.
 \end{eqnarray}
 Understandably that this instability owe its origin to the presence of extraneous charges.}.
 In the next section we consider oscillatory response of such particle which is governed by
 superposition of two restoring forces, namely, the constructive elastic restoring force
 whose strength is dominated by gradient in shear modulus
\begin{eqnarray}
  \label{e3.15}
 f''_i=\nabla_k \sigma_{ik}=2u_{ik}\nabla_k \mu(r)+2\mu\nabla_k\,u_{ik}\approx 2u_{ik}\nabla_k \mu,\quad \quad  \nabla_k\,u_{ik}=\nabla^2 u_i=0.
 \end{eqnarray}
 and the above destructive force of dielectric stresses, $f_i'=\nabla_k\,p_{ik}$ and
 and derive condition of stability which imposes severe constrain on the total charge and radius of the uniformly charged nanoparticle of electret which is of crucial importance to their practically useful applications.

\subsection{Frequencies of electrostriction modes}

 The optically induced vibrations driven by combined the above forces of gradient-shear elastic stress $f''$ and dielectric stresses $f'$ are governed the following equations of electro-elastic dynamics and energy balance
 \begin{eqnarray}
  \label{e4.1}
 &&\rho\,{\ddot u}_i=\nabla_k
 \sigma_{ik}+\nabla_i\,p_{ik}=2u_{ik}\nabla_k \mu+\frac{1}{8\pi}\nabla_k [E_i\,\delta D_k+E_k\,\delta D_i],\quad
 \delta D_i=-4\pi\rho_e\,u_i,\\
 \label{e4.2}
 &&
 \frac{\partial }{\partial t} \int \frac{\rho{\dot u}^2}{2} d{\cal V}=-
 2\int (\nabla_k \mu) u_{ik}{\dot u}_i d{\cal V}
 -\int \rho_e
 [E_i\,u_k+E_k\,u_i]\,{\dot u}_{ik}d{\cal V}
 \end{eqnarray}
 and making use of substitution $u_i({\bf r},t)=a_i({\bf r})\,\alpha(t)$ we obtain
  \begin{eqnarray}
 \label{e4.3}
   && M{\ddot \alpha}+ K\alpha=0,\quad   M = \int \rho\,a_i\,a_i\,d{\cal V},\quad\quad K=K_\mu+K_d, \\
   \label{e4.4}
   && K_\mu = 2 \int\,(\nabla_k \mu)\, a_i\,a_{ik}\,d{\cal V},\quad K_d=-\frac{1}{2}\int \rho_e[E_i\,a_k+E_k\,a_i]
  [\nabla_i\,a_k+\nabla_k\,a_i]\, d{\cal V}.
   \end{eqnarray}
 The mass parameters in spheroidal and torsional modes are given by
  \begin{eqnarray}
  \label{e4.4}
 M^s=4\pi\rho {\cal N}_s^2 R^{2\ell+1}\,\frac{\ell(\ell+1)^2}{(2\ell+1)},\quad M^t=4\pi\rho {\cal N}_t^2 R^{2\ell+3}\, \frac{\ell(\ell+1)}{(2\ell+1)(2\ell+3)}
 \end{eqnarray}
  and for the stiffness one has
 \begin{eqnarray}
 \label{e4.6}
 K^s_\mu = 8\pi {\cal N}_s^2\,\ell(\ell-1)\,\int\limits_{0}^{R} (\nabla_r \mu) \, r^{2\ell-1} dr,\quad  K_d^s=-\frac{32\pi^2}{3}\frac{\rho_e^2}{\epsilon} {\cal N}_s^2 R^{2\ell+1}
 \frac{\ell(\ell+1)(\ell^2-1)}{(2\ell+1)},\\
  \label{e4.7}
 K^t_\mu = 4\pi {\cal N}_t^2\,\frac{\ell(\ell^2-1)}{(2\ell+1)}\,
  \int\limits_{0}^{R} (\nabla_r \mu) \, r^{2\ell+1} dr,\quad K_d^t=
 -\frac{16\pi^2}{3}\frac{\rho_e^2}{\epsilon} {\cal N}_t^2 R^{2\ell+3}\,
 \frac{\ell(\ell^2-1)}{(2\ell+1)(2\ell+3)}.
\end{eqnarray}
These equations provide a basis for computing frequency
$$\omega^2=\frac{K_\mu+K_d}{M}$$
of electrostriction vibrational response of a nanoparticles of uniformly charged electret
with the non-homogeneous shear modulus profile.

As a representative example let us consider a case of nanoparticle with shear modulus profile
given by
\begin{eqnarray}
 \label{e4.7a}
 \mu(r)=\mu\left[1-\left(\frac{r}{R}\right)\right]
 \end{eqnarray}
In this case the frequency of spheroidal electrostriction mode can be conveniently represented in the following form
\begin{eqnarray}
 \nonumber
\omega_s^2&=&\omega_e^2\,2(2\ell+1)(\ell-1)\ell^{-1}-\frac{2}{3}\omega_d^2\,(\ell-1)\\
\label{e4.7}
&=&\omega_e^2\,2(2\ell+1)(\ell-1)\ell^{-1}\left[1-\frac{\ell}{3(2\ell+1)}\beta\right],\quad \beta=\frac{\omega_d^2}{\omega_e^2}
 \end{eqnarray}
and for the torsional electrostriction mode as follows
\begin{eqnarray}
 \nonumber
\omega_t^2&=&\omega_e^2\frac{(2\ell+3)(\ell-1)}{2(\ell+1)}-\frac{1}{3}\omega_d^2\,(\ell-1)\\
\label{e4.8}
&=&\omega_e^2\,\frac{(2\ell+3)(\ell-1)}{2(\ell+1)}\left[1-\frac{2(\ell+1)}{3(2\ell+3)}\beta\right],\quad \beta=\frac{\omega_d^2}{\omega_e^2}
 \end{eqnarray}
 One sees that the lowest overtone of both spheroidal and torsional modes is of quadrupole
 degree $\ell=2$. The absence of monopole $\ell=0$,
 breathing overtone, is the consequence of adapted approximation of
 incompressible matter. The dipole fields of both  poloidal and toroidal displacements
 describe center-of-mass translation and rigid-body rotation, respectively, that is, the
 non-vibrational reaction of nanoparticle.

\vspace*{0.5cm}

 \begin{figure}[ht]
 \centering\
 \includegraphics[width=10.0cm]{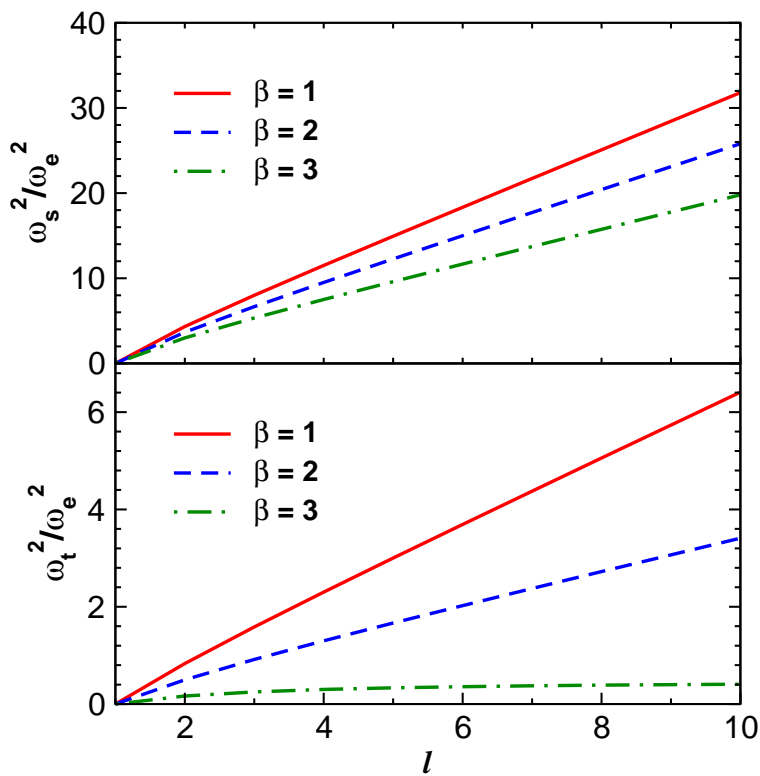}
 \caption{\small Ratio of the squared frequency of spheroidal (upper panel) and torsional (down panel)
 electrostriction modes to the squared frequency of elastic shear oscillations as a function of multipole degree at indicated values of parameter of stability $\beta$.}
 \end{figure}

 In Fig.1 we plot the ratio $\omega_s^2/\omega_e^2$ and $\omega_t^2/\omega_e^2$ as functions of the
 multipole degree of spheroidal and torsional vibrations $\ell$, respectively, showing that the larger the $\ell$ the higher is the frequency.  Also, this figure exhibits strong dependence of frequencies upon the parameter
\begin{eqnarray}
 \label{e4.9}
\beta=\frac{\omega_d^2}{\omega_e^2}=\frac{3}{4\pi}(\epsilon\mu)^{-1}\frac{Q^2}{R^4},\quad Q=\rho_e{\cal V},\quad {\cal V}=(4\pi/3)R^3.
\end{eqnarray}
 carrying information about total charge $Q$ accumulated by particle of radius $R$ and shows
 that the larger $\beta$ (the large the ratio $Q/R^2$) the lower is the frequency.
 The most conspicuous feature of electrostriction response in question is that
 the lowest, quadrupole, overtones become unstable when the parameter $\beta$ attains critical value
 $\beta=\beta_c$. Specifically, the spheroidal electrostriction vibrational mode becomes unstable, meaning  $\omega_s(\ell=2)=0$, when
 \begin{eqnarray}
 \label{e4.10}
 \left[1-\frac{\ell}{3(2\ell+1)}\beta\right]_{\ell=2}=0\quad\to\quad\beta_c^s=\frac{15}{2}.
 \end{eqnarray}
 The lowest quadrupole torsional electrostriction vibrational mode unstable, $\omega_s(\ell=2)=0$, when
\begin{eqnarray}
 \label{e4.12}
\left[1-\frac{2(\ell+1)}{3(2\ell+3)}\beta\right]_{\ell=2}=0\quad\to\quad\beta_c^t=\frac{7}{2}.
\end{eqnarray}

\vspace*{0.5cm}

 \begin{figure}[ht]
 \centering\
 \includegraphics[width=12.0cm]{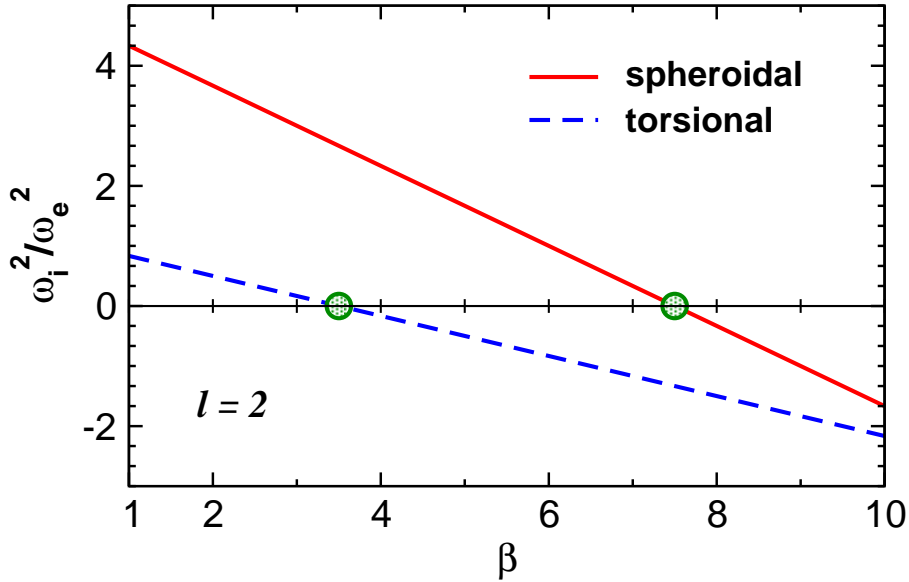}
 \caption{\small The frequency of quadrupole spheroidal and torsional electrostriction modes as a function
 of the stability parameter $\beta$. The circles on 0X-axis mark values of $\beta$ at which the
 electrostriction mode becomes unstable undergoing transition from regime of oscillations to the relaxation regime.}
 \end{figure}

In Fig.2 we plot the ratio of squared frequencies of quadrupole, $\ell=2$, overtones
of both spheroidal and torsion electrostriction modes as functions of the stability parameter $\beta$
by highlighting  the above critical values of stability parameter $\beta_c$ by circles on the 0X-axis.
In this points the electrostriction modes undergo transition from the oscillatory regime to the
relaxation regime. This leads us to conclude that nanoparticle of uniformly charged electret is stable to optically
induced deformation oscillations when
\begin{eqnarray}
 \label{e2.19a}
     \frac{Q}{R^2}<C(\sqrt{\epsilon\mu})
\end{eqnarray}
where constant $C$ falls in the range $5<C<10$; understandably that the
lowest $\beta_c$ should be regarded as genuine critical value of this parameter.
The practical usefulness of the above established conditions of instability is that
it imposes severe constrain on the size and total charge accommodated by the electret nanoparticles
and this must be taken into account in the process for technological fabrication.
As a representative example, for the nano and micro dimensions $10^{-8} < R < 10^{-6}$ m whose total charge is of the order of $Q\approx 10^2 e\approx
10^{-16}$ (in SI units) from polymers with dielectric constant $1 < \epsilon < 15$, and shear modulus
$10^6 < \mu < 10^9$ (in SI units), from above obtaned condition it follows that $10^{-4}< Q/R^2 < 1 $.
So that for ultra fine particles of nano sizes this condition is always fulfilled.

\section{Summary}
  In this, somewhat technical in mathematical details, paper the theory of electrostriction
  response of nanoparticle of non-conducting elastic polymers capable of accumulating an extraneous charge
  to ac electromagnetic load has been developed. The prime purpose was to elucidate stability
  of such nanoparticles to optically induced distortions. This issue is
  of crucial importance to utilization of uniformly charged polymeric ultra fine particles (which are currently produced by jammed technique) as biolabels as well as for other hi-tech application.
  Based on general equations of electro-solid-mechanics appropriate for
  an isotropic electrets capable of accumulating likely-charged inclusions uniformly dispersed over the volume,
  the model of spherical particle of such a material undergoing optically induced
  spheroidal and torsional oscillations restored by superposition of two forces has been considered, namely, the
  constructive elastic force dominated by gradient of shear modulus and destructive force of dielectric stresses has been considered. It is that, contrary to nanoparticle to noble metals (responding to ac field by surface plasmon resonances which are independent upon particle size), the optical response of nanoparticles of
  uniformly charged electrets strongly depends upon their size and total accumulated charge. The conditions of electro-elastic stability of optically induced electrostriction spheroidal and torsional vibrational modes
  has been obtained showing that both modes of such response undergo transition from regime of oscillations
  to the relaxation regime. This condition highlights the most important difference between optically induced
  electrostatic fluctuations in the dielectric nanoparticle of electret which are of volume character
  and electrostatic, plasma, fluctuations in the metal nanoparticle which are of surface character: in this latter case these oscillations are stable whereas in the former case
  the electrostatic fluctuation can lead to disintegration of the electret nanoparticle by means of electric discharge. All these suggest if the condition of electro-elastic stability is fulfilled the optically
  induced electrostriction modes in question can be detected as resonant modes of photo absorbtion and the lowest
  overtones of these modes are of quadrupole degree.

\section{Acknowledgements}

  This work is partly supported by NSC of Taiwan, under grants NSC-
  97-2811-M-007-003.

\end{document}